\documentclass[twocolumn, times] {aastex631}
\usepackage{amsmath}
\usepackage{amssymb}
\usepackage{nicefrac}
\usepackage{booktabs}

\graphicspath{{./}{figures/}}

\hypersetup{citecolor=blue}
\begin{document}

%==================================================================================================================================================================================================================================================================
\title{NECOLA: Towards a Universal Field-level Cosmological Emulator}

\correspondingauthor{Neerav Kaushal}
\email{kaushal@mtu.edu}

\author[0000-0003-4786-2348]{Neerav Kaushal}
\affiliation{Department of Physics, Michigan Technological University, Houghton, MI 49931, USA}

\author[0000-0002-4816-0455]{Francisco Villaescusa-Navarro}
\affiliation{Center for Computational Astrophysics, 162 5th Avenue, New York, NY 10010, USA}
\affiliation{Department of Astrophysical Sciences, Princeton University, Peyton Hall, Princeton NJ 08544, USA}

\author[0000-0003-3052-3059]{Elena Giusarma}
\affiliation{Department of Physics, Michigan Technological University, Houghton, MI 49931, USA}

\author[0000-0002-0701-1410]{Yin Li}
\affiliation{Center for Computational Astrophysics, 162 5th Avenue, New York, NY 10010, USA}
\affiliation{Center for Computational Mathematics, 162 5th Avenue, New York, NY 10010, USA}

\author{Conner Hawry}
\affiliation{Department of Physics, Michigan Technological University, Houghton, MI 49931, USA}

\author{Mauricio Reyes}
\affiliation{Department of Physics, Michigan Technological University, Houghton, MI 49931, USA}

%==================================================================================================================================================================================================================================================================
\begin{abstract}

We train convolutional neural networks to correct the output of fast and approximate N-body simulations at the field level. Our model, Neural Enhanced COLA --NECOLA--, takes as input a snapshot generated by the computationally efficient COLA code and corrects the positions of the cold dark matter particles to match the results of full N-body Quijote simulations. We quantify the accuracy of the network using several summary statistics, and find that NECOLA can reproduce the results of the full N-body simulations with sub-percent accuracy down to $k\simeq1~h{\rm Mpc}^{-1}$. Furthermore, the model, that was trained on simulations with a fixed value of the cosmological parameters, is also able to correct the output of COLA simulations with different values of $\Omega_{\rm m}$, $\Omega_{\rm b}$, $h$, $n_s$, $\sigma_8$, $w$, and $M_\nu$ with very high accuracy: the power spectrum and the cross-correlation coefficients are within $\simeq1\%$ down to $k=1~h{\rm Mpc}^{-1}$. Our results indicate that the correction to the power spectrum from fast/approximate simulations or field-level perturbation theory is rather universal. Our model represents a first step towards the development of a fast field-level emulator to sample not only primordial mode amplitudes and phases, but also the parameter space defined by the values of the cosmological parameters.

\end{abstract}


%==================================================================================================================================================================================================================================================================
\section{Introduction} \label{sec:intro}

In order to extract valuable information about fundamental physics from cosmic surveys, we need theoretical predictions to confront the collected data. On semi-linear scales, analytic tools like perturbation theory \citep{Bernardeau_review} can be used to provide such theoretical predictions. However, on non-linear scales, where a large amount of cosmological information resides \citep[e.g.][]{Quijote, Samushia_2021, Gualdi_2021, Kuruvilla_2021, Bayer_2021,  Banerjee_2019, Changhoon_2019, Uhlemann_2020, Friedrich_2020, Massara_2020, Dai_2020, Allys_2020, Banerjee_2020, Banerjee_2021, Gualdi_2020, Gualdi_2021, Giri_2020, Bella_2020, Changhoon_2020, Valgiannis_2021, Bayer_2021,  Kuruvilla_2021b}, numerical simulations become necessary. 

Cosmological simulations can be classified into two broad categories: 1) N-body simulations that model the matter field accounting only for the force of gravity, and 2) hydrodynamic simulations that model not only gravity but also fluid hydrodynamics and astrophysical effects such as the formation of stars and feedback from black holes. While computationally more efficient than hydrodynamic simulations, N-body simulations are still expensive, and running large sets or high-resolution simulations require a significant computational cost \citep[e.g.][]{Quijote, Abacus, uchuu, bacco, Aemulus, Aemulus2, Aemulus3, Aemulus4, AbacusSummit, DarkQuest}. To overcome this, several methods have been developed that are much less computationally demanding but come at the expense of being less accurate (e.g., ALPT \citep{ALPT}, PThalos \citep{PTHALOS}, PINOCCHIO \citep{pinocchio}, FastPM \citep{FastPM}, COLA \citep{cola,cola2, L-PICOLA},  EZMOCKS \citep{EZmocks}, FlowPM \citep{flowPM}, PATCHY \citep{PATCHY}, log-normal models \citep{lognormal,lognormal2}, HALOGEN \citep{HALOGEN}, QPM \citep{QPM}, HaloNet \citep{halonet} and mass-Peak Patch \citep{peak-patch, peak_patch}).

% \Neerav{ADDED MORE CITATIONS}
% \Neerav{ADDED CITATIONS} 

Being able to run fast and accurate simulations is of main importance in cosmology in order to provide the theoretical predictions needed to retrieve the maximum information from cosmological surveys. In this work, we try to build a bridge between the fast and approximate COLA simulations, and the expensive and accurate full N-body simulations using deep learning. We build on the work of \cite{Siyu_2018} and \cite{Renan_2020} who used neural networks to find the mapping between the displacement field generated by the Zel'dovich approximation to the one from fast and full N-body simulations, respectively. In this work, we train convolutional neural networks to correct the particle positions from COLA simulation snapshots to match those of full N-body Quijote simulations. The most important conclusion of our work is that our model seems to be universal, i.e., once trained on simulations with a fixed value of cosmological parameters, our network is able to correct the particle positions of COLA simulations with any other cosmology with surprising accuracy: the power spectrum is accurate at the 1\% level down to $k=1~h{\rm Mpc}^{-1}$.

This paper is organized as follows. In Section \ref{sec:methods}, we describe the simulations we use and the architecture of our neural network model. We present the results of the trained network in Section \ref{sec:results}. Finally, we draw our conclusions in Section \ref{sec:conclusions}.

%==================================================================================================================================================================================================================================================================
\section{Methods} 
\label{sec:methods}

In this section, we describe the two types of simulations we used, together with the model architecture and the training procedure.

\subsection{Simulations}

\subsubsection{Full N-body simulations}
We made use of the Quijote full N-body simulations \citep{Quijote} to both train and test the model. The simulations used in this work follow the evolution of $512^3$ cold dark matter (CDM) particles (plus $512^3$ neutrino particles in the case of massive neutrino cosmologies) from $z=127$ down to $z=0$ in a periodic volume of $(1000~h^{-1}{\rm Mpc})^3$. We train the network using a set of 100 simulations from the fiducial cosmology, where the values of the cosmological parameters are fixed to: $\Omega_{\rm m}=0.3175$, $\Omega_{\rm b}=0.049$, $h=0.6711$, $n_s=0.9624$, $\sigma_8=0.834$, $w=-1$, $M_\nu=0.0$ eV. These simulations are only different in the value of the initial random seed.

We test the accuracy of our network on simulations with very different cosmologies to the one used in the training. For this, we made use of 100 of 2,000 simulations of the latin hypercube contained in the Quijote simulations, where the values of the cosmological parameters span the range $\Omega_{\rm m}\in[0.1,0.5]$, $\Omega_{\rm b}\in[0.03,0.07]$, $h\in[0.5,0.9]$, $n_s\in[0.8,1.2]$ and $\sigma_8\in[0.6,1.0]$. In these simulations, not only the set of values of the cosmological parameters is different, but the initial random seed varies as well. Furthermore, we test the accuracy of our network on models with massive neutrinos and on models where the dark energy equation of state is $w\neq-1$, making use of Quijote simulations labeled $M_\nu^+$, $M_\nu^{++}$, $M_\nu^{+++}$, $w^{+}$, and $w^{-1}$. On average, each of the Quijote simulations used in this work required $\sim500$ CPU hours to run. We refer the reader to \cite{Quijote} for further details on the Quijote simulations.

\subsubsection{Approximate N-body simulations}

The fast and approximate simulations we use in this work are run with the COmoving Lagrangian Acceleration (COLA) \citep[][]{cola} method, that combines second-order Lagrangian perturbation theory (2LPT) \citep[][]{Bernardeau_2002} on large scales with N-body methods on small scales. In particular, we use the MG-PICOLA \citep[][]{mgpicola} package. For each Quijote full N-body simulation, we run a COLA simulation by matching 1) the number of particles, 2) the set of values of the cosmological parameters, and 3) the value of the initial random seed, which gives rise to identical initial Gaussian field for both Quijote and COLA. These simulations require fewer time steps than the full N-body simulations and are therefore much more computationally efficient. Each COLA simulation is run with $30$ time steps equally spaced in log from $z=9$ down to $z=0$. On average, these simulations only take $3$ CPU hours to run.

\subsection{Model}

\subsubsection{Input and Target}
Let us write the displacement vector of a particle as $\vec{d} = \vec{x}_f - \vec{x}_i$, where $\vec{x}_f$ and $\vec{x}_i$ are the final ($z=0$) and initial (Lagrangian) position of the particle. Our goal is to train a neural network to correct the positions of the particles generated by COLA, to match them with those from a full N-body simulation, i.e.
\begin{equation}
\vec{x}_{f,{\rm Nbody}}=g(\vec{x}_{f,{\rm COLA}}) 
\label{eq1}
\end{equation}
where $g$ is an unknown function. Note that the right-hand side of Eq. \ref{eq1} should not be taken as the position of the particular particle considered, but also of all its neighboring particles. To preserve translational equivariance, we use displacement vectors instead of absolute particle positions. Thus, the input to the network is  $\vec{d}_{\rm COLA}$, rather than $\vec{x}_{f,{\rm COLA}}$. The network is trained to learn $\vec{d}_{\rm Nbody}-\vec{d}_{\rm COLA}=\vec{x}_{f,{\rm Nbody}}-\vec{x}_{f,{\rm COLA}}$.% and predict $\vec{d}_{\rm Nbody}$.

\subsubsection{Model}

We follow \citet{deoliveira2020fast} and use a V-Net \citep[][]{milletari2016vnet} based model that consists of $2$ downsampling and $2$ upsampling layers connected in a "V" shape. Blocks of two $3^3$ convolutions connect the input, the resampling, and the output layers. $1^3$ convolutions are added over each of these convolution blocks to realize a residual connection. We add batch normalization after every convolution except the first one and the last two, and leaky ReLU activation with a negative slope of $0.01$ after every batch normalization, as well as the first and the second to last convolutions. The last activation in each residual block acts after the summation, following \citet{milletari2016vnet}. As in U-Net/V-Net, at all resolution levels (with the exception of the bottleneck levels), the inputs to the downsampling layers are concatenated to the outputs of the upsampling layers. All layers have a channel size of $64$, except for the input and the output, that have $3$ channels (the displacement vector along each cartesian coordinate), as well as those after concatenations ($128$-channeled). Finally, the input $(\vec{d}_{\rm COLA})$ is directly added to the output, so that the network could learn the corrections to match the target $(\vec{d}_{\rm Nbody}-\vec{d}_{\rm COLA})$. Stride-2 $2^3$ convolutions and stride-$\nicefrac{1}{2}$ $2^3$ transposed convolutions are used in downsampling and upsampling layers, respectively.

Following \citet{deoliveira2020fast}, we minimize a loss function given by $L=\rm log_{e}(L_{\delta} L_{\Delta}^{\lambda})$, where $L_{\delta}$ is the Mean Squared Error (MSE) loss on $n(\textbf{x})$ (the particle number in voxel $\textbf{x}$) and $L_{\Delta}$ is the MSE on the displacement vector $\vec{d}$.
With this loss function, we are able to train the model to make accurate predictions in both Lagrangian and Eulerian spaces. By combining the two losses with logarithm rather than summation, we can account for their absolute magnitudes and trade between their relative values. $\lambda$ here serves as a weight on this trade-off of relative losses and $\lambda=1$ works pretty well in our case.

The input cannot be fed into the network at once due to the big size of the data $(3 \times 512^3)$, and we thus divide it into smaller chunks first. We crop the data into subcubes of size $3 \times 128^3$, corresponding to a simulation box of length $250~h^{-1}{\rm Mpc}$. In order to preserve the physical translational equivariance, no padding has been used in the $3^3$ convolutions, which results in an output that is smaller than the input in spatial size. This limitation is compensated by padding the input cubes periodically with $20$ voxels on each side so that the effective spatial size of the input becomes $3 \times 168^3$. Furthermore, data augmentation is implemented to enforce the equivariance of displacement fields under rotational and parity transformations. We use the Adam optimizer \citep[][]{kingma2017adam} with a learning rate of $0.0001$, $\beta_1 = 0.9$ and $\beta_2 = 0.999$, and reduce the learning rate by half when the loss does not improve for $3$ epochs. The model is trained on $70$ realizations for $100$ epochs and the remaining realizations are used for validation ($20$) and final testing ($10$). From now on, we will refer to this model as NECOLA, from Neural Enhanced COLA, in order to avoid any confusion with the model by \citet{deoliveira2020fast}, which uses Zel'dovich simulations as input and a different value of $\lambda$. Note that the model architecture of NECOLA is the same as that of \citet{deoliveira2020fast}.

\subsection{Benchmark models}
%\Neerav{ADDED BENCHMARK}
In order to compare the predictions of our model, we have used three different benchmarks:
\begin{itemize}
\item \textbf{COLA}. This benchmark represents the results of running the COLA simulation itself.
\item \textbf{ZA}. In this case, the positions of the particles at $z=0$ are computed using the Zel'dovich approximation.
\item \textbf{NN(ZA)}. This benchmark is the model developed by \citet{Renan_2020}, that takes as input ZA simulations and corrects the output to match full N-body simulations. We refer the reader to \citet{Renan_2020} for further details on this model.
\end{itemize}

%%%=================================================================================================================================================================================
\section{Results} \label{sec:results}

In this section, we investigate the performance of our model. We first make use of several summary statistics to quantify the accuracy of our model for simulations with the same cosmology as the one used to train the network. Then, we investigate how well does our network extrapolate to other cosmological models.

\subsection{Fiducial cosmology}

We first present the results of testing the network on simulations that have the same cosmology as the one used for its training.

\subsubsection{Visual comparison}

\begin{figure*}[htb!]
\begin{center}
    %\vspace*{-1.5cm}
    %\hspace*{0.7cm}
	\includegraphics[scale=0.82]{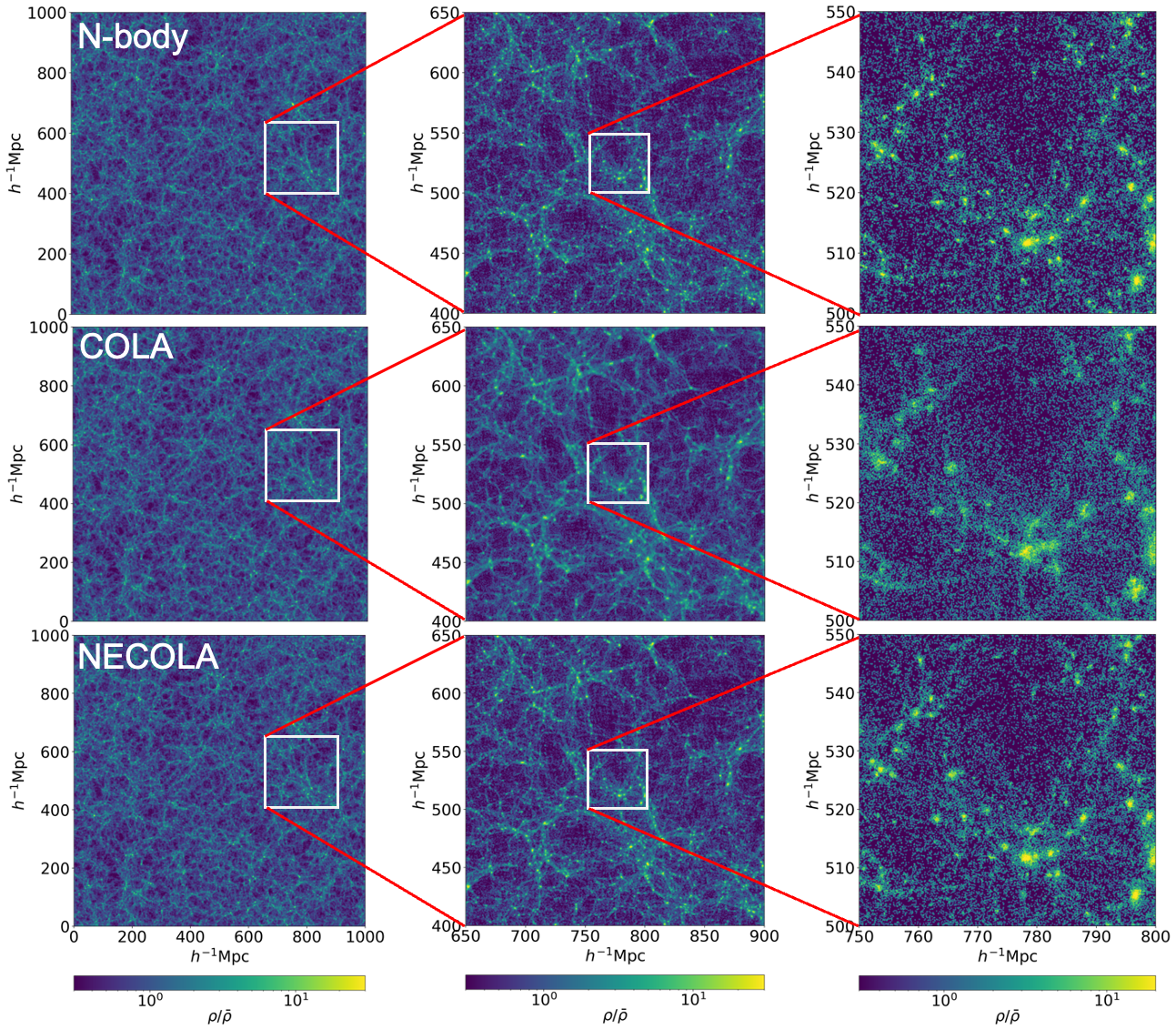}
    \caption{The figure shows the cold dark matter density fields for the target N-body simulations (top), the input/benchmark COLA simulations (middle) and the predictions of our model (bottom), at a scale of $1000\; \text{Mpc h}^{-1}$ (left column), $250\; \text{Mpc h}^{-1}$ (middle column) and $50 \; \text{Mpc h}^{-1}$ (right column). Each figure is a zoomed-in image of the white box in the figure on its left.
    }
    \label{fig2}
    \vspace{15pt}
    \end{center}
\end{figure*}

Before quantifying the accuracy of the network using summary statistics, we perform a visual inspection of its output. In Fig. \ref{fig2}, we show the distribution of matter at $z=0$ from the full N-body simulation (top row), the COLA simulation (middle row), and NECOLA (bottom row). 

While looking at large scales, the agreement between the three methods is really good, but when we look at small scales, some differences are visible. In the case of COLA, the output is more diffuse and halos do not exhibit a high concentration in their centers, in contrast to the corresponding N-body simulation. On the other hand, NECOLA produces much sharper results, clearly defining the positions and boundaries of dark matter halos. 

\subsubsection{Power spectrum}

\begin{figure*}[htb!]
    %\vspace*{-1.5cm}
    \hspace*{0.1cm}
	\includegraphics[scale=0.49]{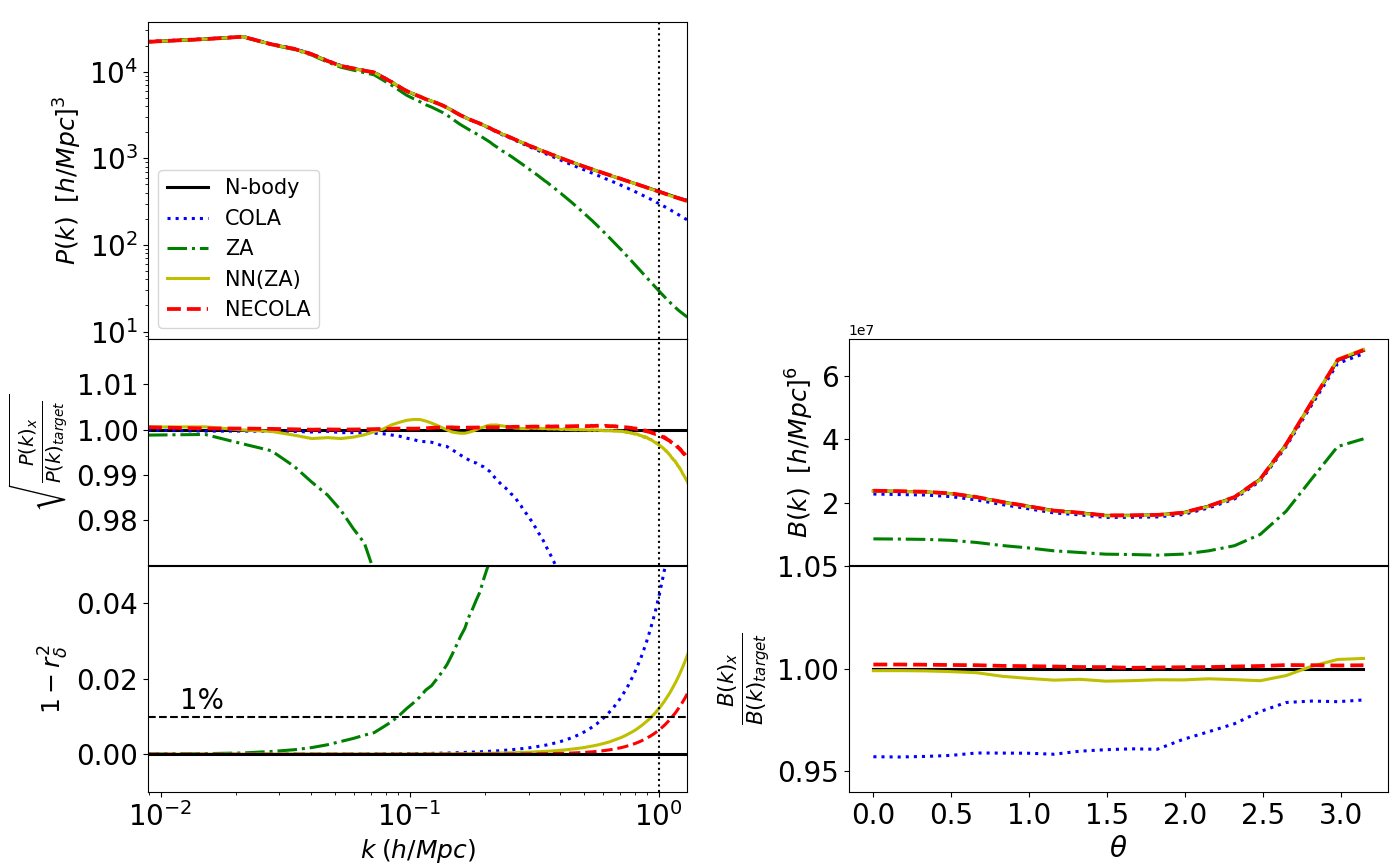}
    \caption{The left plot shows the 3D matter power spectrum (top), the transfer function (middle) and the cross-correlation coefficient (bottom), while the right plot shows the bispectrum for $k_1=0.15~h{\rm Mpc}^{-1}$ and $k_2=0.25~h{\rm Mpc}^{-1}$ (top) and the bispectrum ratio (bottom) for the target N-body simulations (solid black), the COLA simulations (dotted blue), the ZA approximations (dash-dotted green), NN(ZA) (solid yellow), and NECOLA (dashed red). As can be seen, NECOLA outperforms all benchmarks in all cases.
    }
    \label{fig1}
\end{figure*}

The power spectrum is defined as the Fourier transform of the 2-point correlation function (2PCF), which measures the excess probability of finding a pair of random galaxies (or points) at a given separation compared to the one from a random distribution. The power spectrum is one of the most important summary statistics used in cosmology since for Gaussian density fields (like the one our Universe resembles on large, linear scales), it fully characterizes the statistical properties of the field.

In the top-left panel of Fig. \ref{fig1}, we show with a solid black line the average power spectra from $10$ Quijote simulations of the test set. The dotted blue line shows the average power spectrum from the corresponding COLA simulations, while the green dot-dashed line outputs the average power spectrum of Zel'dovich-evolved simulations. The solid yellow and dashed red lines show the average power spectrum from NN(ZA) and NECOLA, respectively. As can be seen, the worst model is the one that only employs the Zel'dovich approximation, followed by the COLA simulation.

In order to better visualize the differences between the output of the N-body simulation and the networks, we plot in the middle-left panel of Fig. \ref{fig1} the transfer function, defined as
\begin{equation}
    T(k) = \sqrt{\frac{P_{\text{pred}}(k)}{P_{\text{target}}(k)}},
\end{equation}
where $P_{\text{pred}}(k)$ and $P_{\text{target}}(k)$ are the average matter power spectra of the predictions and the target density fields respectively. Values close to 1 indicate a better agreement between the prediction and the target. As can be seen, both networks achieve a sub-percent accuracy on the power spectrum down to $k=1~h{\rm Mpc}^{-1}$, though the results obtained from NECOLA are slightly more accurate. We note that in the case of the Quijote simulations, it does not make sense to look into much smaller scales than $k~\sim1~h{\rm Mpc}^{-1}$, as those are not numerically converged in the simulations due to mass resolution \citep{Quijote}.

\subsubsection{Cross-Correlation Coefficient}
In Fourier space, every mode can be written as $\delta(\vec{k})=Ae^{i\theta}$, where $A$ and $\theta$ are the mode amplitude and phase, respectively. When using the power spectrum, we are effectively comparing how well the amplitude of the modes from the network and the simulation agree. However, that statistic neglects the correlations in mode phases, which are very important in the non-linear regime. To quantify the correlations between the mode phases, we use the cross-correlation coefficient, $r$, defined as
\begin{equation}
    r(k) = \frac  {P_{\text{pred}\times\text{target}}(k)}  {\sqrt{P_{\text{pred}}(k)P_{\text{target}}(k)}},
\end{equation}
where the numerator is the cross-power spectrum between the predictions and the target and the denominator contains the auto-power spectrum of the prediction and the target. Values of $r$ close to 1 indicate a very good correlation in mode phases. In the bottom-left panel of Fig. \ref{fig1}, we show the cross-correlation coefficient averaged over the testing set for the different cases considered. We find that NECOLA achieves the highest accuracy, being within 1\% down to $k=1~h{\rm Mpc}^{-1}$.

\subsubsection{Bispectrum}

The third statistic that we consider to quantify the agreement between the full simulations and the network predictions is the bispectrum, defined as 
\begin{equation}
    \langle\delta_{\mathbf{k_1}} \delta_{\mathbf{k_2}} \delta_{\mathbf{k_3}}\rangle \equiv {\delta_{D}(\mathbf{k_{123}})} B(\mathbf{k_1},\mathbf{k_2},\mathbf{k_3})  ,
\end{equation}
where $\delta(\mathbf{k})$ the overdensity in the Fourier space and $\mathbf{k_{123}}$ $\equiv \mathbf{k}_1 + \mathbf{k}_2 + \mathbf{k}_3$.

Differently to the power spectrum, the bispectrum quantifies the correlation between triplets of modes in closed triangles. For Gaussian density fields, this quantity is zero, and therefore, its amplitude and shape capture information about the non-Gaussianities in a given field. In the top-right panel of Fig. \ref{fig1}, we show the bispectrum for $k_1=0.15~h{\rm Mpc}^{-1}$ and $k_2=0.25~h{\rm Mpc}^{-1}$ as a function of the angle between $k_1$ and $k_2$, $\theta$. On this scale, we cannot see large differences, besides the fact that the Zel'dovich approximation underestimates the amplitude of the bispectrum, as expected. In the bottom-right panel of Fig. \ref{fig1}, we show the ratio between the different bispectra to the bispectrum of the N-body simulation. We find that both neural networks give very accurate results, although NECOLA is slightly more accurate.

\begin{figure*}[htb!]
    \hspace*{-0.4cm}
	\includegraphics[scale=0.55]{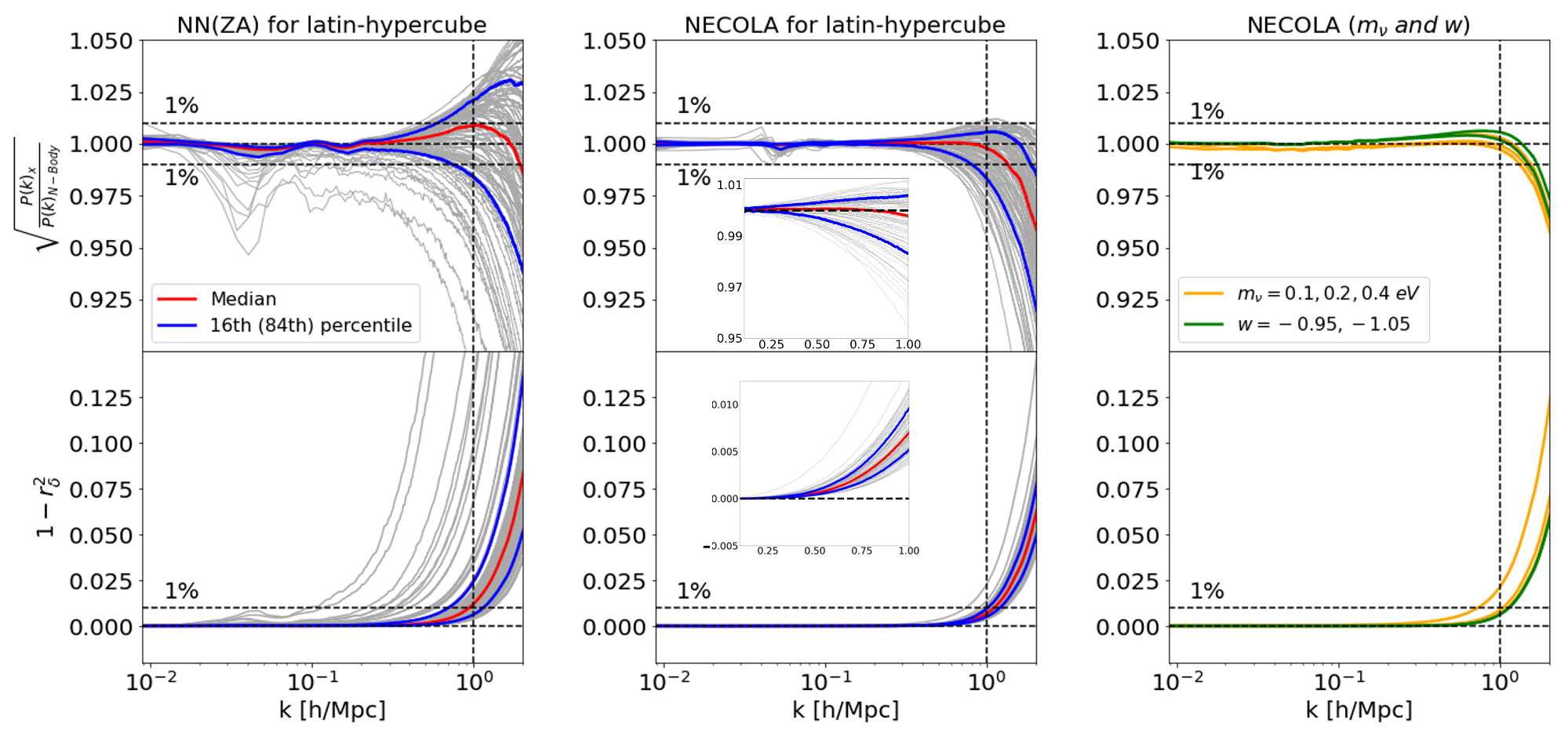}
    \caption{We test the NN(ZA) and NECOLA models, which are trained on simulations with a fixed cosmology, on models with very different values of the cosmological parameters. The left and middle panels show the results when using 100 simulations of the Quijote latin-hypercube (that vary $\Omega_{\rm m}$, $\Omega_{\rm b}$, $h$, $n_s$, and $\sigma_8$), while the right panel displays the results for cosmologies with massive neutrinos and a dark energy equation of state different to $-1$. The red lines represent the median while the blue lines represent the 16th (and 84th) percentile of the predictions. As can be seen, NECOLA not only performs better than NN(ZA), but it is surprisingly accurate all the way down to $k\sim1~h{\rm Mpc}^{-1}$. Besides, it also works for models with massive neutrinos and $w\neq-1$. The curve with the largest difference in the neutrino cross-correlation coefficient corresponds to a model with $M_\nu=0.4$ eV.}
    \vspace{15pt}
    \label{fig3}
\end{figure*}

\subsection{Model Extrapolation}

We now explore how our model extrapolates to cosmologies different from the one used to train the model.

We first test the extrapolation properties of the model on the parameters $\Omega_{\rm m}$, $\Omega_{\rm b}$, $h$, $n_s$, and $\sigma_8$ by using 100 simulations of the Quijote latin-hypercube set. We emphasize that for the simulations in this set, the values of these 5 cosmological parameters are varied at the same time, together with the value of the initial random seed. For each of these simulations, we run its COLA counterpart and input it to the network, which corrects the positions of the particles. 

For each cosmology, we compute the power spectrum of the output of NECOLA and of the full N-body Quijote simulation. In Fig. \ref{fig3}, we show in the middle panel the transfer function together with the cross-correlation coefficient. As can be seen, NECOLA is able to correct the output of the COLA simulations in all cases with surprising accuracy: below $\simeq1\%$ down to $k=1~h{\rm Mpc}^{-1}$. 

Next, we repeat the same exercise but using NN(ZA) and show the results in the left panel of Fig. \ref{fig3}. As can be seen, the network trained on COLA snapshots exhibits much stronger extrapolation features than the one trained on Zel'dovich displacements.

We now investigate if NECOLA is also able to correct COLA outputs for simulations with massive neutrinos. We emphasize that no simulations used for training the model contain massive neutrinos. For this, we made use of simulations from the $M_\nu^+$, $M_\nu^{++}$, and $M_\nu^{+++}$ Quijote sets, corresponding to cosmologies with sums of the neutrino masses equal to 0.1 eV, 0.2 eV, and 0.4 eV. In these simulations, we have both dark matter and neutrino particles. From each set, we take 10 simulations and run their COLA counterpart. Next, we input to NECOLA the displacement vectors of the dark matter particles of the COLA simulation, and NECOLA outputs the corrected positions of the dark matter particles for these massive neutrino models.

\begin{table*}[htb!]
%\vspace*{-0.4cm}
  \caption{Computational cost associated to running a full N-body simulation, a COLA simulation, and NECOLA. Note that in case of NECOLA, we report the GPU wall time.}
  \label{tab1}
  \centering
  \begin{tabular}{llll}
    \toprule
    Simulation & N-body (QUIJOTE) & Fast (COLA) & NECOLA (PyTorch-GPU) \\
    \midrule
    CPU-/GPU-sec & $10^6$ & $10^4$ & $125$ \\
    \bottomrule
    \vspace{10pt}
  \end{tabular}
\end{table*}

In the right panel of Fig. \ref{fig3}, we show the results of this calculation with yellow lines. As can be seen, NECOLA is able to correct the positions of the dark matter particles such that their power spectrum and cross-correlation coefficient agree with the full N-body calculation below $1\%$ down to $k=1~h{\rm Mpc}^{-1}$. We note that although our network only works with the cold dark matter field, assuming a linear neutrino field correlated with the initial Gaussian field will, for most of the cases, give very accurate predictions for the total matter field \citep{Massara_2014}. On the other hand, the cold dark matter field is the one responsible for the abundance and clustering of dark matter halos and galaxies \citep{Paco2014, Ema2014}. It is interesting to note that NECOLA also gives more accurate results than the model by \citet{giusarma2019learning}, that used neural networks to map from N-body simulations without neutrinos to N-body simulations with neutrinos.

Lastly, we study the performance of our model for cosmologies with values of the dark energy equation of state, $w$, different from -1. For this, we made use of the 10 simulations of the $w^+$ and $w^-$ Quijote sets, that have a value of $w$ equal to-0.95 and -1.05, respectively. For each of these simulations, we run their COLA counterpart and compute the displacement vectors. We then input those into the network that returns the corrected positions of the dark matter particles. In the right panel of Fig. \ref{fig3}, we show with green lines the results of computing the transfer function and cross-correlation coefficient between the output of the network and the full N-body simulations. As can be seen, in this case as well, NECOLA is able to correct the output of the cosmologies that it has never seen before.

\subsection{Computational cost}
A typical N-body simulation takes roughly $500$ CPU hours to run, or $\sim 10^6$ CPU seconds, while a single COLA simulation takes around $3$ CPU hours or $\sim 10^4$ CPU seconds. We run our CNN model on $1$ GPU ($320$ NVIDIA P$100$-$16$GB) using PyTorch \citep[][]{NEURIPS2019_bdbca288} and it takes $\sim125$ GPU seconds to run. A runtime comparison of the target, benchmark, and our model is shown in Table \ref{tab1}. Thus, in practice, the main limitation of our model comes from the computational cost associated with running COLA simulations itself. Despite this, our model allows us to speed up the computational cost by a factor of 100.

\section{Summary} 
\label{sec:conclusions}

Providing accurate theoretical predictions is necessary in order to extract the maximum amount of information from upcoming cosmological surveys. The computational cost of running full N-body simulations is currently too expensive to carry out standard analysis such as MCMC. On the other hand, fast simulations can reduce the computational cost by orders of magnitude at the expense of sacrificing accuracy. 

In this work, we have shown that we can use neural networks to correct the output of approximate simulations to match full N-body simulations from the Quijote suite. Our model, coined NECOLA, from Neural Enhanced COLA, has been trained on simulations with a fixed value of the cosmological parameters. We have shown that our model is not only able to correct the output of COLA simulations run with the same cosmology as the one used to train the network, but is also able to correct COLA simulations that have very different values of the parameters $\Omega_{\rm m}$, $\Omega_{\rm b}$, $h$, $n_s$, $\sigma_8$, $M_\nu$, and $w$. This surprising feature of our network indicates that the correction from the output of COLA to a full N-body might be universal, i.e. independent of cosmology.

This may have important consequences for perturbation theory studies, that are able to accurately model the linear and perturbative regime but fail on non-linear scales. Our work indicates that a generic, cosmology-independent correction may be feasible, at least in the case of the power spectrum.

Our network can be used as a field-level emulator to explore not only the initial modes amplitudes and phases \citep[][]{borg1}, but also the cosmological parameter space \citep[][]{borg2}. We note however that further work is needed to claim that our model is precise for statistics other than the power spectrum and cross-correlation function when using it in extrapolation. In future work, we will quantify the accuracy of our network on other summary statistics like bispectrum, halo mass function, etc. Additional work is also needed to incorporate velocities into this framework, that will allow performing studies in redshift-space. Besides, further work is needed to quantify the accuracy of NECOLA at redshifts other than the one used for training, together with the universality of the network under changes of simulation resolution.

Overall, this work opens an interesting direction in the development of fast and generalized field-level emulators needed to maximize the scientific return of upcoming cosmological missions.

The trained models, predictions and statistics extracted from the testing and extrapolation sets are hosted under the public github repository \url{https://github.com/neeravkaushal/cola-to-nbody.git} and the model has been trained using the map2map code \url{https://github.com/eelregit/map2map.git}.

\section{Acknowledgments} \label{sec:ack}
The Quijote simulations used in this work are publicly available at \url{https://github.com/franciscovillaescusa/Quijote-simulations.git}. The COLA simulations have been run using the MG-PICOLA code, publicly available at  \url{https://github.com/HAWinther/MG-PICOLA-PUBLIC.git}. We acknowledge that our work has been performed using the Princeton Research Computing resources at Princeton University which is a consortium of groups led by the Princeton Institute for Computational Science and Engineering (PICSciE). E.G. thanks Michigan Space Grant Consortium for their support. The work of FVN has been supported by the WFIRST program through NNG26PJ30C and NNN12AA01C. FVN and YL are supported by the Simons Foundation. Research reported in this publication was supported in part by funding provided by the National Aeronautics and Space Administration (NASA), under award number NNX15AJ20H, Michigan Space Grant Consortium (MSGC).

\bibliography{sample631}{}

\begin{thebibliography}{}
\expandafter\ifx\csname natexlab\endcsname\relax\def\natexlab#1{#1}\fi
\providecommand{\url}[1]{\href{#1}{#1}}
\providecommand{\dodoi}[1]{doi:~\href{http://doi.org/#1}{\nolinkurl{#1}}}
\providecommand{\doeprint}[1]{\href{http://ascl.net/#1}{\nolinkurl{http://ascl.net/#1}}}
\providecommand{\doarXiv}[1]{\href{https://arxiv.org/abs/#1}{\nolinkurl{https://arxiv.org/abs/#1}}}

\bibitem[{Agrawal {et~al.}(2017)Agrawal, Makiya, Chiang, Jeong, Saito, \&
  Komatsu}]{lognormal2}
Agrawal, A., Makiya, R., Chiang, C.-T., {et~al.} 2017, Journal of Cosmology and
  Astroparticle Physics, 2017, 003–003, \dodoi{10.1088/1475-7516/2017/10/003}

\bibitem[{{Allys} {et~al.}(2020){Allys}, {Marchand}, {Cardoso},
  {Villaescusa-Navarro}, {Ho}, \& {Mallat}}]{Allys_2020}
{Allys}, E., {Marchand}, T., {Cardoso}, J.~F., {et~al.} 2020, \prd, 102,
  103506, \dodoi{10.1103/PhysRevD.102.103506}

\bibitem[{{Alves de Oliveira} {et~al.}(2020{\natexlab{a}}){Alves de Oliveira},
  {Li}, {Villaescusa-Navarro}, {Ho}, \& {Spergel}}]{Renan_2020}
{Alves de Oliveira}, R., {Li}, Y., {Villaescusa-Navarro}, F., {Ho}, S., \&
  {Spergel}, D.~N. 2020{\natexlab{a}}, arXiv e-prints, arXiv:2012.00240.
\newblock \doarXiv{2012.00240}

\bibitem[{{Alves de Oliveira} {et~al.}(2020{\natexlab{b}}){Alves de Oliveira},
  {Li}, {Villaescusa-Navarro}, {Ho}, \& {Spergel}}]{deoliveira2020fast}
---. 2020{\natexlab{b}}, arXiv e-prints, arXiv:2012.00240.
\newblock \doarXiv{2012.00240}

\bibitem[{Angulo {et~al.}(2021)Angulo, Zennaro, Contreras, Aricò,
  Pellejero-Ibañez, \& Stücker}]{bacco}
Angulo, R.~E., Zennaro, M., Contreras, S., {et~al.} 2021, Monthly Notices of
  the Royal Astronomical Society, 507, 5869–5881,
  \dodoi{10.1093/mnras/stab2018}

\bibitem[{Avila {et~al.}(2015)Avila, Murray, Knebe, Power, Robotham, \&
  Garcia-Bellido}]{HALOGEN}
Avila, S., Murray, S.~G., Knebe, A., {et~al.} 2015, Monthly Notices of the
  Royal Astronomical Society, 450, 1856–1867, \dodoi{10.1093/mnras/stv711}

\bibitem[{{Banerjee} \& {Abel}(2021{\natexlab{a}})}]{Banerjee_2020}
{Banerjee}, A., \& {Abel}, T. 2021{\natexlab{a}}, \mnras, 500, 5479,
  \dodoi{10.1093/mnras/staa3604}

\bibitem[{{Banerjee} \& {Abel}(2021{\natexlab{b}})}]{Banerjee_2021}
---. 2021{\natexlab{b}}, \mnras, 504, 2911, \dodoi{10.1093/mnras/stab961}

\bibitem[{{Banerjee} {et~al.}(2020){Banerjee}, {Castorina},
  {Villaescusa-Navarro}, {Court}, \& {Viel}}]{Banerjee_2019}
{Banerjee}, A., {Castorina}, E., {Villaescusa-Navarro}, F., {Court}, T., \&
  {Viel}, M. 2020, \jcap, 2020, 032, \dodoi{10.1088/1475-7516/2020/06/032}

\bibitem[{Bayer {et~al.}(2021)Bayer, Villaescusa-Navarro, Massara, Liu,
  Spergel, Verde, Wandelt, Viel, \& Ho}]{Bayer_2021}
Bayer, A.~E., Villaescusa-Navarro, F., Massara, E., {et~al.} 2021, Detecting
  neutrino mass by combining matter clustering, halos, and voids.
\newblock \doarXiv{2102.05049}

\bibitem[{Berger \& Stein(2018)}]{halonet}
Berger, P., \& Stein, G. 2018, Monthly Notices of the Royal Astronomical
  Society, 482, 2861–2871, \dodoi{10.1093/mnras/sty2949}

\bibitem[{{Bernardeau} {et~al.}(2002{\natexlab{a}}){Bernardeau}, {Colombi},
  {Gazta{\~n}aga}, \& {Scoccimarro}}]{Bernardeau_review}
{Bernardeau}, F., {Colombi}, S., {Gazta{\~n}aga}, E., \& {Scoccimarro}, R.
  2002{\natexlab{a}}, \physrep, 367, 1, \dodoi{10.1016/S0370-1573(02)00135-7}

\bibitem[{{Bernardeau} {et~al.}(2002{\natexlab{b}}){Bernardeau}, {Colombi},
  {Gazta{\~n}aga}, \& {Scoccimarro}}]{Bernardeau_2002}
---. 2002{\natexlab{b}}, \physrep, 367, 1,
  \dodoi{10.1016/S0370-1573(02)00135-7}

\bibitem[{{Bond} \& {Myers}(1996)}]{peak-patch}
{Bond}, J.~R., \& {Myers}, S.~T. 1996, \apjs, 103, 1, \dodoi{10.1086/192267}

\bibitem[{{Castorina} {et~al.}(2014){Castorina}, {Sefusatti}, {Sheth},
  {Villaescusa-Navarro}, \& {Viel}}]{Ema2014}
{Castorina}, E., {Sefusatti}, E., {Sheth}, R.~K., {Villaescusa-Navarro}, F., \&
  {Viel}, M. 2014, \jcap, 2, 049, \dodoi{10.1088/1475-7516/2014/02/049}

\bibitem[{{Chuang} {et~al.}(2015){Chuang}, {Kitaura}, {Prada}, {Zhao}, \&
  {Yepes}}]{EZmocks}
{Chuang}, C.-H., {Kitaura}, F.-S., {Prada}, F., {Zhao}, C., \& {Yepes}, G.
  2015, \mnras, 446, 2621, \dodoi{10.1093/mnras/stu2301}

\bibitem[{{Coles} \& {Jones}(1991)}]{lognormal}
{Coles}, P., \& {Jones}, B. 1991, \mnras, 248, 1, \dodoi{10.1093/mnras/248.1.1}

\bibitem[{{Dai} {et~al.}(2020){Dai}, {Verde}, \& {Xia}}]{Dai_2020}
{Dai}, J.-P., {Verde}, L., \& {Xia}, J.-Q. 2020, \jcap, 2020, 007,
  \dodoi{10.1088/1475-7516/2020/08/007}

\bibitem[{{de la Bella} {et~al.}(2020){de la Bella}, {Tessore}, \&
  {Bridle}}]{Bella_2020}
{de la Bella}, L.~F., {Tessore}, N., \& {Bridle}, S. 2020, arXiv e-prints,
  arXiv:2011.06185.
\newblock \doarXiv{2011.06185}

\bibitem[{{DeRose} {et~al.}(2019){DeRose}, {Wechsler}, {Tinker}, {Becker},
  {Mao}, {McClintock}, {McLaughlin}, {Rozo}, \& {Zhai}}]{Aemulus}
{DeRose}, J., {Wechsler}, R.~H., {Tinker}, J.~L., {et~al.} 2019, \apj, 875, 69,
  \dodoi{10.3847/1538-4357/ab1085}

\bibitem[{{Feng} {et~al.}(2016){Feng}, {Chu}, \& {Seljak}}]{FastPM}
{Feng}, Y., {Chu}, M.-Y., \& {Seljak}, U. 2016, ArXiv e-prints.
\newblock \doarXiv{1603.00476}

\bibitem[{{Friedrich} {et~al.}(2020){Friedrich}, {Uhlemann},
  {Villaescusa-Navarro}, {Baldauf}, {Manera}, \& {Nishimichi}}]{Friedrich_2020}
{Friedrich}, O., {Uhlemann}, C., {Villaescusa-Navarro}, F., {et~al.} 2020,
  \mnras, 498, 464, \dodoi{10.1093/mnras/staa2160}

\bibitem[{Garrison {et~al.}(2018)Garrison, Eisenstein, Ferrer, Tinker, Pinto,
  \& Weinberg}]{Abacus}
Garrison, L.~H., Eisenstein, D.~J., Ferrer, D., {et~al.} 2018, Astrophys. J.
  Suppl., 236, 43, \dodoi{10.3847/1538-4365/aabfd3}

\bibitem[{{Giri} \& {Smith}(2020)}]{Giri_2020}
{Giri}, U., \& {Smith}, K.~M. 2020, arXiv e-prints, arXiv:2010.07193.
\newblock \doarXiv{2010.07193}

\bibitem[{{Giusarma} {et~al.}(2019){Giusarma}, {Reyes Hurtado},
  {Villaescusa-Navarro}, {He}, {Ho}, \& {Hahn}}]{giusarma2019learning}
{Giusarma}, E., {Reyes Hurtado}, M., {Villaescusa-Navarro}, F., {et~al.} 2019,
  arXiv e-prints, arXiv:1910.04255.
\newblock \doarXiv{1910.04255}

\bibitem[{{Gualdi} {et~al.}(2021{\natexlab{a}}){Gualdi}, {Gil-Marin}, \&
  {Verde}}]{Gualdi_2021}
{Gualdi}, D., {Gil-Marin}, H., \& {Verde}, L. 2021{\natexlab{a}}, arXiv
  e-prints, arXiv:2104.03976.
\newblock \doarXiv{2104.03976}

\bibitem[{{Gualdi} {et~al.}(2021{\natexlab{b}}){Gualdi}, {Novell},
  {Gil-Mar{\'\i}n}, \& {Verde}}]{Gualdi_2020}
{Gualdi}, D., {Novell}, S., {Gil-Mar{\'\i}n}, H., \& {Verde}, L.
  2021{\natexlab{b}}, \jcap, 2021, 015, \dodoi{10.1088/1475-7516/2021/01/015}

\bibitem[{{Hahn} \& {Villaescusa-Navarro}(2021)}]{Changhoon_2020}
{Hahn}, C., \& {Villaescusa-Navarro}, F. 2021, \jcap, 2021, 029,
  \dodoi{10.1088/1475-7516/2021/04/029}

\bibitem[{{Hahn} {et~al.}(2020){Hahn}, {Villaescusa-Navarro}, {Castorina}, \&
  {Scoccimarro}}]{Changhoon_2019}
{Hahn}, C., {Villaescusa-Navarro}, F., {Castorina}, E., \& {Scoccimarro}, R.
  2020, \jcap, 2020, 040, \dodoi{10.1088/1475-7516/2020/03/040}

\bibitem[{{He} {et~al.}(2019){He}, {Li}, {Feng}, {Ho}, {Ravanbakhsh}, {Chen},
  \& {P{\'o}czos}}]{Siyu_2018}
{He}, S., {Li}, Y., {Feng}, Y., {et~al.} 2019, Proceedings of the National
  Academy of Science, 116, 13825, \dodoi{10.1073/pnas.1821458116}

\bibitem[{{Howlett} {et~al.}(2015){Howlett}, {Manera}, \&
  {Percival}}]{L-PICOLA}
{Howlett}, C., {Manera}, M., \& {Percival}, W.~J. 2015, Astronomy and
  Computing, 12, 109, \dodoi{10.1016/j.ascom.2015.07.003}

\bibitem[{Ishiyama {et~al.}(2021)Ishiyama, Prada, Klypin, Sinha, Metcalf,
  Jullo, Altieri, Cora, Croton, de~la Torre, \& et~al.}]{uchuu}
Ishiyama, T., Prada, F., Klypin, A.~A., {et~al.} 2021, Monthly Notices of the
  Royal Astronomical Society, 506, 4210–4231, \dodoi{10.1093/mnras/stab1755}

\bibitem[{{Jasche} \& {Wandelt}(2013)}]{borg1}
{Jasche}, J., \& {Wandelt}, B.~D. 2013, \apj, 779, 15,
  \dodoi{10.1088/0004-637X/779/1/15}

\bibitem[{Jasche \& Wandelt(2013)}]{borg2}
Jasche, J., \& Wandelt, B.~D. 2013, Monthly Notices of the Royal Astronomical
  Society, 432, 894, \dodoi{10.1093/mnras/stt449}

\bibitem[{Kingma \& Ba(2017)}]{kingma2017adam}
Kingma, D.~P., \& Ba, J. 2017, Adam: A Method for Stochastic Optimization.
\newblock \doarXiv{1412.6980}

\bibitem[{Kitaura \& Heß(2013)}]{ALPT}
Kitaura, F.-S., \& Heß, S. 2013, Monthly Notices of the Royal Astronomical
  Society: Letters, 435, L78, \dodoi{10.1093/mnrasl/slt101}

\bibitem[{{Kitaura} {et~al.}(2014){Kitaura}, {Yepes}, \& {Prada}}]{PATCHY}
{Kitaura}, F.-S., {Yepes}, G., \& {Prada}, F. 2014, \mnras, 439, L21,
  \dodoi{10.1093/mnrasl/slt172}

\bibitem[{{Kuruvilla}(2021)}]{Kuruvilla_2021b}
{Kuruvilla}, J. 2021, arXiv e-prints, arXiv:2109.13938.
\newblock \doarXiv{2109.13938}

\bibitem[{{Kuruvilla} \& {Aghanim}(2021)}]{Kuruvilla_2021}
{Kuruvilla}, J., \& {Aghanim}, N. 2021, arXiv e-prints, arXiv:2102.06709.
\newblock \doarXiv{2102.06709}

\bibitem[{{Maksimova} {et~al.}(2021){Maksimova}, {Garrison}, {Eisenstein},
  {Hadzhiyska}, {Bose}, \& {Satterthwaite}}]{AbacusSummit}
{Maksimova}, N.~A., {Garrison}, L.~H., {Eisenstein}, D.~J., {et~al.} 2021,
  \mnras, \dodoi{10.1093/mnras/stab2484}

\bibitem[{{Massara} {et~al.}(2021){Massara}, {Villaescusa-Navarro}, {Ho},
  {Dalal}, \& {Spergel}}]{Massara_2020}
{Massara}, E., {Villaescusa-Navarro}, F., {Ho}, S., {Dalal}, N., \& {Spergel},
  D.~N. 2021, \prl, 126, 011301, \dodoi{10.1103/PhysRevLett.126.011301}

\bibitem[{{Massara} {et~al.}(2014){Massara}, {Villaescusa-Navarro}, \&
  {Viel}}]{Massara_2014}
{Massara}, E., {Villaescusa-Navarro}, F., \& {Viel}, M. 2014, \jcap, 2014, 053,
  \dodoi{10.1088/1475-7516/2014/12/053}

\bibitem[{McClintock {et~al.}(2019{\natexlab{a}})McClintock, Rozo, Becker,
  DeRose, Mao, McLaughlin, Tinker, Wechsler, \& Zhai}]{Aemulus2}
McClintock, T., Rozo, E., Becker, M.~R., {et~al.} 2019{\natexlab{a}},
  Astrophys. J., 872, 53, \dodoi{10.3847/1538-4357/aaf568}

\bibitem[{McClintock {et~al.}(2019{\natexlab{b}})McClintock, Rozo, Banerjee,
  Becker, DeRose, McLaughlin, Tinker, Wechsler, \& Zhai}]{Aemulus4}
McClintock, T., Rozo, E., Banerjee, A., {et~al.} 2019{\natexlab{b}}, The
  Aemulus Project IV: Emulating Halo Bias.
\newblock \doarXiv{1907.13167}

\bibitem[{Milletari {et~al.}(2016)Milletari, Navab, \&
  Ahmadi}]{milletari2016vnet}
Milletari, F., Navab, N., \& Ahmadi, S.-A. 2016, V-Net: Fully Convolutional
  Neural Networks for Volumetric Medical Image Segmentation.
\newblock \doarXiv{1606.04797}

\bibitem[{{Modi} {et~al.}(2020){Modi}, {Lanusse}, \& {Seljak}}]{flowPM}
{Modi}, C., {Lanusse}, F., \& {Seljak}, U. 2020, arXiv e-prints,
  arXiv:2010.11847.
\newblock \doarXiv{2010.11847}

\bibitem[{Monaco {et~al.}(2002)Monaco, Theuns, \& Taffoni}]{pinocchio}
Monaco, P., Theuns, T., \& Taffoni, G. 2002, Monthly Notices of the Royal
  Astronomical Society, 331, 587, \dodoi{10.1046/j.1365-8711.2002.05162.x}

\bibitem[{{Nishimichi} {et~al.}(2019){Nishimichi}, {Takada}, {Takahashi},
  {Osato}, {Shirasaki}, {Oogi}, {Miyatake}, {Oguri}, {Murata}, {Kobayashi}, \&
  {Yoshida}}]{DarkQuest}
{Nishimichi}, T., {Takada}, M., {Takahashi}, R., {et~al.} 2019, \apj, 884, 29,
  \dodoi{10.3847/1538-4357/ab3719}

\bibitem[{Paszke {et~al.}(2019)Paszke, Gross, Massa, Lerer, Bradbury, Chanan,
  Killeen, Lin, Gimelshein, Antiga, Desmaison, Kopf, Yang, DeVito, Raison,
  Tejani, Chilamkurthy, Steiner, Fang, Bai, \& Chintala}]{NEURIPS2019_bdbca288}
Paszke, A., Gross, S., Massa, F., {et~al.} 2019, in Advances in Neural
  Information Processing Systems, ed. H.~Wallach, H.~Larochelle,
  A.~Beygelzimer, F.~d\textquotesingle Alch\'{e}-Buc, E.~Fox, \& R.~Garnett,
  Vol.~32 (Curran Associates, Inc.).
\newblock
  \url{https://proceedings.neurips.cc/paper/2019/file/bdbca288fee7f92f2bfa9f7012727740-Paper.pdf}

\bibitem[{{Samushia} {et~al.}(2021){Samushia}, {Slepian}, \&
  {Villaescusa-Navarro}}]{Samushia_2021}
{Samushia}, L., {Slepian}, Z., \& {Villaescusa-Navarro}, F. 2021, arXiv
  e-prints, arXiv:2102.01696.
\newblock \doarXiv{2102.01696}

\bibitem[{{Scoccimarro} \& {Sheth}(2002)}]{PTHALOS}
{Scoccimarro}, R., \& {Sheth}, R.~K. 2002, \mnras, 329, 629,
  \dodoi{10.1046/j.1365-8711.2002.04999.x}

\bibitem[{{Stein} {et~al.}(2019){Stein}, {Alvarez}, \& {Bond}}]{peak_patch}
{Stein}, G., {Alvarez}, M.~A., \& {Bond}, J.~R. 2019, \mnras, 483, 2236,
  \dodoi{10.1093/mnras/sty3226}

\bibitem[{Tassev {et~al.}(2015)Tassev, Eisenstein, Wandelt, \&
  Zaldarriaga}]{cola2}
Tassev, S., Eisenstein, D.~J., Wandelt, B.~D., \& Zaldarriaga, M. 2015, sCOLA:
  The N-body COLA Method Extended to the Spatial Domain.
\newblock \doarXiv{1502.07751}

\bibitem[{Tassev {et~al.}(2013)Tassev, Zaldarriaga, \& Eisenstein}]{cola}
Tassev, S., Zaldarriaga, M., \& Eisenstein, D.~J. 2013, Journal of Cosmology
  and Astroparticle Physics, 2013, 036, \dodoi{10.1088/1475-7516/2013/06/036}

\bibitem[{{Uhlemann} {et~al.}(2020){Uhlemann}, {Friedrich},
  {Villaescusa-Navarro}, {Banerjee}, \& {Codis}}]{Uhlemann_2020}
{Uhlemann}, C., {Friedrich}, O., {Villaescusa-Navarro}, F., {Banerjee}, A., \&
  {Codis}, S. 2020, \mnras, 495, 4006, \dodoi{10.1093/mnras/staa1155}

\bibitem[{{Valogiannis} \& {Dvorkin}(2021)}]{Valgiannis_2021}
{Valogiannis}, G., \& {Dvorkin}, C. 2021, arXiv e-prints, arXiv:2108.07821.
\newblock \doarXiv{2108.07821}

\bibitem[{{Villaescusa-Navarro} {et~al.}(2014){Villaescusa-Navarro}, {Marulli},
  {Viel}, {Branchini}, {Castorina}, {Sefusatti}, \& {Saito}}]{Paco2014}
{Villaescusa-Navarro}, F., {Marulli}, F., {Viel}, M., {et~al.} 2014, \jcap, 3,
  011, \dodoi{10.1088/1475-7516/2014/03/011}

\bibitem[{{Villaescusa-Navarro} {et~al.}(2020){Villaescusa-Navarro}, {Hahn},
  {Massara}, {Banerjee}, {Delgado}, {Ramanah}, {Charnock}, {Giusarma}, {Li},
  {Allys}, {Brochard}, {Uhlemann}, {Chiang}, {He}, {Pisani}, {Obuljen}, {Feng},
  {Castorina}, {Contardo}, {Kreisch}, {Nicola}, {Alsing}, {Scoccimarro},
  {Verde}, {Viel}, {Ho}, {Mallat}, {Wandelt}, \& {Spergel}}]{Quijote}
{Villaescusa-Navarro}, F., {Hahn}, C., {Massara}, E., {et~al.} 2020, \apjs,
  250, 2, \dodoi{10.3847/1538-4365/ab9d82}

\bibitem[{White {et~al.}(2013)White, Tinker, \& McBride}]{QPM}
White, M., Tinker, J.~L., \& McBride, C.~K. 2013, Monthly Notices of the Royal
  Astronomical Society, 437, 2594, \dodoi{10.1093/mnras/stt2071}

\bibitem[{{Wright} {et~al.}(2017){Wright}, {Winther}, \& {Koyama}}]{mgpicola}
{Wright}, B.~S., {Winther}, H.~A., \& {Koyama}, K. 2017, \jcap, 2017, 054,
  \dodoi{10.1088/1475-7516/2017/10/054}

\bibitem[{Zhai {et~al.}(2019)Zhai, Tinker, Becker, DeRose, Mao, McClintock,
  McLaughlin, Rozo, \& Wechsler}]{Aemulus3}
Zhai, Z., Tinker, J.~L., Becker, M.~R., {et~al.} 2019, Astrophys. J., 874, 95,
  \dodoi{10.3847/1538-4357/ab0d7b}

\end{thebibliography}
\bibliographystyle{aasjournal}

%% This command is needed to show the entire author+affiliation list when
%% the collaboration and author truncation commands are used.  It has to
%% go at the end of the manuscript.
%\allauthors

%% Include this line if you are using the \added, \replaced, \deleted
%% commands to see a summary list of all changes at the end of the article.
%\listofchanges

\end{document}